\documentclass[showpacs,aps,pre,twocolumn]{revtex4}
\usepackage{graphicx}

\usepackage[usenames]{color}
\usepackage{cancel}
\usepackage{hyperref}
\usepackage[normalem]{ulem}

\begin{document}

\title{Witness of unsatisfiability for a random $3$-satisfiability formula}

\author{Lu-Lu Wu and
Hai-Jun Zhou}
\affiliation{State Key Laboratory for Theoretical Physics, Institute
of Theoretical Physics, Chinese Academy of Sciences, Beijing 100190, China}

\author{Mikko Alava}
\affiliation{Department of Applied Physics,
Aalto University,  FI-00076 Aalto, Finland}

\author{Erik Aurell}
\affiliation{ACCESS Linnaeus Center,
KTH, Sweden; Department of Computational Biology, AlbaNova
University Center, 10691 Stockholm, Sweden;
Aalto University School of Science, FI-00076 Aalto, Finland}

\author{Pekka Orponen}
\affiliation{Department of Information and Computer Science, Aalto University,
FI-00076 Aalto, Finland}

\date{\today}

\begin{abstract}
The random $3$-satisfiability ($3$-SAT) problem is in the
unsatisfiable (UNSAT) phase when the clause density $\alpha$ exceeds
a critical value $\alpha_s \approx  4.267$. However, rigorously proving the
unsatisfiability of a given
large  $3$-SAT instance is extremely difficult.
In this paper we apply the mean-field theory of statistical physics to the
unsatisfiability problem, and
show that a specific type of UNSAT witnesses (Feige-Kim-Ofek witnesses)
can in principle be constructed
when the clause density $\alpha > 19$. We then construct
Feige-Kim-Ofek witnesses
for single $3$-SAT instances through a simple
random sampling algorithm  and a focused local search algorithm.
The random sampling algorithm works only when $\alpha$ scales at least linearly
 with the variable number $N$, but the focused local search algorithm
works for clause densty $\alpha > c N^{b}$ with $b \approx 0.59$ and
prefactor $c \approx 8$.
The exponent $b$ can be further decreased by enlarging the single
parameter $S$ of the focused local search algorithm.
\end{abstract}

\pacs{89.70.Eg, 89.20.Ff, 02.10.Ox, 75.10.Nr}

\maketitle

\section{Introduction}

The satisfiability (SAT) problem is a constraint satisfaction problem
of great practical and theoretical importance. On the practical side, many
constraint satisfaction problems and combinatorial optimization problems in
industry and engineering can be converted into a SAT problem, therefore
many heuristic solution-searching algorithms have been developed over the years
 for single problem instances (see review \cite{Gomes-etal-2008}).
On the theoretical side, the SAT problem is the
first constraint satisfaction problem shown to be NP-complete
\cite{Cook-1971,Garey-Johnson-1979}, all other NP-complete problems
can be transformed into the SAT problem through a polynomial number of steps.
Understanding the computational complexity of the SAT problem has attracted a
lot of research efforts.

The ensemble
of random $K$-SAT problem has been the focus of intensive theoretical
studies by computer scientists and statistical physicists in the last
twenty years
\cite{Cheeseman-Kanefsky-Taylor-1991,Mitchell-etal-1992,Kirkpatrick-Selman-1994,Monasson-Zecchina-1996,Mezard-etal-2002,Mezard-Zecchina-2002,Krzakala-etal-PNAS-2007,Alava-etal-2008}.
In a given instance (formula) of the random $K$-SAT problem,
the states of $N$ binary variables are constrained by $M$ clauses, with each
clause involving a fixed number $K$ of variables, randomly and independently
chosen from the whole set of $N$ variables. The clause density is defined as
$$
\alpha \equiv \frac{M}{N} \; ,
$$
which is just the ratio between the clause number $M$ and the variable number
$N$.

The random $K$-SAT problem has a critical clause density $\alpha_s(K)$ at
which a satisfiability transition occurs.
At the thermodynamic limit of $N\rightarrow \infty$, all the $M$
clauses of an instance of the
random $K$-SAT problem can be simultaneously satisfied if the clause density
$\alpha < \alpha_s(K)$, but this becomes
impossible if $\alpha > \alpha_s(K)$. The value of $\alpha_s(K)$ for $K\geq 3$
can be estimated by the mean-field theory of statistical physics
\cite{Mezard-etal-2002,Mezard-Zecchina-2002,Mertens-etal-2006}.
For example $\alpha_s(3)=4.267$ for the random $3$-SAT problem.

Most previous investigations on the random $K$-SAT problem considered the
SAT phase, $\alpha < \alpha_s(K)$.
To prove a $K$-SAT formula is satisfiable, it is
sufficient to show that there exists a single  spin configuration of
the $N$ variables which makes all the $M$ clauses to be
simultaneously satisfied.
However, to certify a $K$-SAT formula to be unsatisfiable is much harder.
In principle one has to
show that none of the $2^N$ spin configurations
satisfies the $M$ clauses simultaneously.

Theoretical computer scientists
have approached the $K$-SAT problem from the UNSAT phase through
spectral algorithms
\cite{Goerdt-Krivelevich-2001,Feige-Ofek-2004,CojaOghlan-Goerdt-Lanka-2007}. These refutation algorithms are able to certify the unsatisfiability of
random $3$-SAT formulas when $\alpha > c N^{\frac{1}{2}}$ (where the
constant $c$ should be sufficiently large). The refutation lower-bound
for random $3$-SAT
was further pushed to $\alpha > c N^{\frac{2}{5}}$ by Feige, Kim and
Ofek \cite{Feige-Kim-Ofek-2006} from another theoretical approach,
namely treating  a given $3$-SAT instance also as a $3$-exclusive-or
($3$-XORSAT) instance. Feige and co-authors \cite{Feige-Kim-Ofek-2006}
observed that, if a $3$-SAT formula is satisfiable, the ground-state energy
of the same formula treated as a $3$-XORSAT can not exceed certain value.
Then proving the unsatisfiability of a $3$-SAT instance is converted to
constructing a high-enough lower-bound for the corresponding $3$-XORSAT
ground-state energy. The method of Ref.~\cite{Feige-Kim-Ofek-2006} therefore
gives an indirect witness that there is no configuration which can
simultaneously satisfy all the $M$ clauses of the $3$-SAT instance.
In this paper we refer to such witnesses as Feige-Kim-Ofek (FKO) witnesses.

We study the unsatisfiability of the random $3$-SAT problem both
theoretically and algorithmically in this paper.
 The theoretical question we ask  is: Do Feige-Kim-Ofek
witnesses exist in random $3$-SAT formulas
with large but constant clause density $\alpha$? We give a positive
answer to this question by using (non-rigorous) mean-field method of
statistical physics. We show that FKO witnesses are presented in
large random $3$-SAT formulas provided their
clause density $\alpha > 19$. But
constructing FKO witnesses for such sparse formulas is expected to be
very difficult. A very simple random sampling algorithm is
tested
in this paper. Without any optimization, the performance of this
naive algorithm is not good, it only works for $\alpha$ scaling at least
linearly with $N$.
We then test the performance of a simple focused local search
algorithm. We find this algorithm performs much better, it
can construct UNSAT witnesses for $3$-SAT instances with
clause density $\alpha > 8 N^{0.59}$.
Further improvements  are observed when some modifications
are made on this focused local search algorithm.

The paper is structured as follows: in Sec.~\ref{sec:FKO}
we review the main ideas behind FKO witnesses;
 Sec.~\ref{sec:exist} demonstrates the existence
of FKO witnesses for the sparse random $3$-SAT problem
and Sec.~\ref{sec:sample} shows the performances of the
naive random sampling algorithm and the focused local
search algorithm.
In Sect.~\ref{sec:further} we conclude and discuss further directions
of this work.

\section{The Feige-Kim-Ofek witness}
\label{sec:FKO}

Consider a system with $N$ variables $i\in \{1, 2,\ldots, N\}$. Each
variable $i$ has a (binary) spin state $\sigma_i \in \{-1, +1\}$.
A configuration of the  system is denoted as $\underline{\sigma}
\equiv (\sigma_1, \sigma_2, \ldots, \sigma_N)$,
there are a total number $2^N$ of
such configurations. The system has also
$M$ clauses $a\in \{1, 2, \ldots, M\}$.
Each clause $a$ is a constraint over $K=3$ different variables
(say $i, j, k$), it has three binary coupling constants
(say $J_a^i, J_a^j, J_a^k$),
each of which is either $+1$ or $-1$. We consider two types of energies
for clause $a$, namely the SAT energy
\begin{equation}
E_a^{{\rm sat}}(\sigma_i, \sigma_j, \sigma_k)
= \frac{(1-J_a^i \sigma_i) (1-J_a^j \sigma_j)
(1-J_a^k \sigma_k)}{8} \; ,
\end{equation}
and the XORSAT energy
\begin{equation}
E_a^{{\rm xor}}(\sigma_i, \sigma_j, \sigma_k) =
\frac{1- J_a^i J_a^j J_a^k \sigma_i \sigma_j \sigma_k}{2} \; .
\end{equation}

If the total energy of the system is defined as the sum of all the
SAT energies, then the problem is a $3$-SAT formula with energy
function
\begin{equation}
\label{eq:eSAT}
E^{{\rm sat}}(\underline{\sigma}) = \sum\limits_{a=1}^{M} E_a^{{\rm sat}} \; .
\end{equation}
A configuration $\underline{\sigma}$ is referred to as a satisfying
assignment (or a solution) for the $3$-SAT formula if its energy
$E^{{\rm sat}}(\underline{\sigma}) = 0$. The $3$-SAT formula is
referred to as satisfiable
(SAT) if there exists at least one satisfying assignment
for this formula, otherwise it is referred to as
unsatisfiable (UNSAT).

For the same set of $M$ clauses, we can also consider all the XORSAT
energies and define a $3$-XORSAT formula with energy function
\begin{equation}
\label{eq:eXORSAT}
E^{{\rm xor}}(\underline{\sigma}) = \sum\limits_{a=1}^{M} E_a^{{\rm xor}} \; .
\end{equation}
The ground-state (minimum) energy of the XORSAT energy is denoted as
$E_0^{{\rm xor}}$, namely
$$
E_0^{{\rm xor}} \equiv \min\limits_{\underline{\sigma}} E^{{\rm xor}}(\underline{\sigma})
\; .
$$
Checking whether a $3$-XORSAT formula is satisfiable (namely
$E_0^{{\rm xor}}=0$) is an easy computational tast (it can
be solved by Gaussian elimination).
However if $E_0^{{\rm xor}}>0$, to determine the precise value of $E_0^{{\rm xor}}$ is
a NP-hard computational problem.

\begin{figure}
\begin{center}
\includegraphics[width=0.6\linewidth]{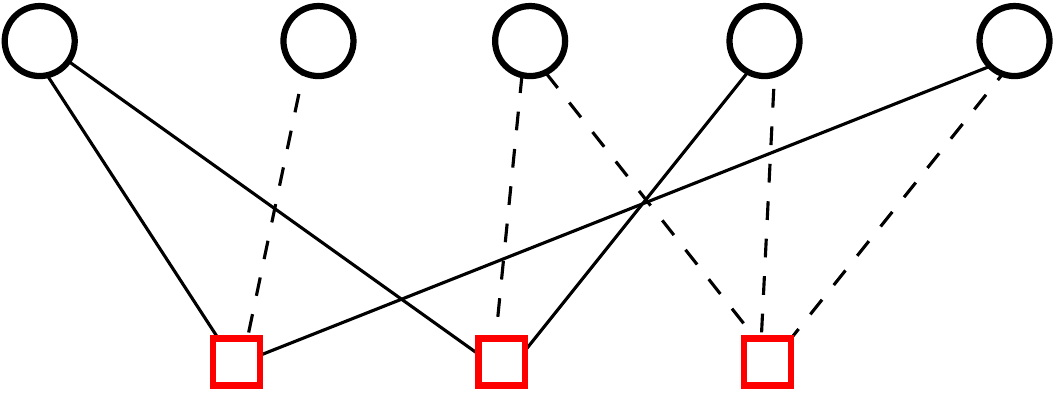}
\end{center}
\caption{
\label{fig:factorgraph}
(color online).
factor graph representation for a $3$-SAT formula. The variables and
clauses are represented by circles and squares, respectively.
Each clause has $3$ edges attached. A solid edge between a variable $i$ and
a clause $a$ means that the coupling constant $J_a^i=1$, while a
dashed edge means that $J_a^i=-1$.
}
\end{figure}

The  constrained system can be
conveniently represented as a bipartite graph with $N$ circular nodes
for the variables and $M$ square nodes for the constraint clauses and
$3 M$ edges between the variable nodes and the clause nodes,
see Fig.~\ref{fig:factorgraph} \cite{Kschischang-etal-2001}.
Such a bipartite graph is referred to as a $3$-SAT factor graph
in this paper.
In the factor graph, each clause $a$ is connected by $3$ edges to
the $3$ constrained variables, and the edge $(i, a)$ between a variable $i$
and a clause $a$ is shown as a solid line (if $J_a^i=1$) or a
dashed line (if $J_a^i=-1$).
In the factor graph of the system, the number of attached edges of
different variables might be different. For a variable $i$ the number
of attached positive and negative edges is denoted as $k_i^+$ and $k_i^-$,
respectively.

To prove the unsatisfiability of a $3$-SAT formula is very
challenging. In principle one has to show that for each of
the $2^N$ configurations, the SAT energy $E^{{\rm sat}}(\underline{\sigma}) > 0$,
but such an enumeration becomes impossible for systems with $N > 1000$.
Feige, Kim, and Ofek (FKO) \cite{Feige-Kim-Ofek-2006}
approached this problem with the proposal of constructing UNSAT witnesses
through the $3$-XORSAT energy (\ref{eq:eXORSAT}).
Here we review their main
ideas \cite{Feige-Kim-Ofek-2006}.

Consider a given $3$-SAT formula with energy function (\ref{eq:eSAT}).
Suppose
this formula is satisfiable, then there is at least one satisfying
configuration $\underline{\sigma}$ such that
$E^{{\rm sat}}(\underline{\sigma})=0$. An edge $(i, a)$ is referred to as being
satisfied by $\underline{\sigma}$ if (and only if) the spin of variable
$i$ is $\sigma_i = J_a^i$ in this configuration.
With respect to $\underline{\sigma}$, the
total number of clauses containing one, two, and three satisfied edges is
denoted as $M_1$, $M_2$ and $M_3$, respectively.
These three integers satisfy the following two relations:
\begin{eqnarray}
 M_1 + M_2 + M_3 & = & M  \; , \label{eq:satm1} \\
 M_1 + 2 M_2 + 3 M_3 & \leq & \frac{3 M}{2} + \frac{1}{2} \sum\limits_{i=1}^{N}
|k_i^+ - k_i^-|  \; . \label{eq:satm2}
\end{eqnarray}
Equation~(\ref{eq:satm1}) is a consequence of the assumption that
$E^{{\rm sat}}(\underline{\sigma}) = 0$, while Eq.~(\ref{eq:satm2}) is due to the
fact that each variable $i$ in its spin state $\sigma_i$ can satisfy at most
$\max(k_i^+, k_i^-)$ edges. The above two expressions lead to
\begin{equation}
\label{eq:m2}
M_2 \leq
 2 M_{1 2} - \frac{3}{2} M  + \frac{1}{2} \sum\limits_{i=1}^{N}
|k_i^+ - k_i^-| \; ,
\label{eq20110520-01}
\end{equation}
where $M_{1 2} \equiv M_1 + M_2$.

On the other hand, it is very easy to check that the $3$-XORSAT
energy (\ref{eq:eXORSAT})
of the configuration $\underline{\sigma}$ is just
$E^{{\rm xor}}(\underline{\sigma}) = M_2$. Therefore,
if $E^{{\rm sat}}(\underline{\sigma})=0$,
then the $3$-XORSAT ground-state energy $E_0^{{\rm xor}}$ must
not exceed $M_2$.
If $E_0^{{\rm xor}}$ exceeds $M_2$ then the $3$-SAT energy
function (\ref{eq:eSAT}) must be positive for all
the $2^N$ configurations. A high-enough $3$-XORSAT ground-state
energy then serves as a FKO witness that the corresponding $3$-SAT
formula is UNSAT.

Consider any spin configuration $\underline{\sigma}$
(not necessarily a configuration with $E^{{\rm sat}}(\underline{\sigma})=0$),
the value of $M_{1 2}$ in Eq.~(\ref{eq20110520-01})
is calculated as
\begin{eqnarray}
M_{1 2} &=& \sum\limits_{a=1}^M \frac{(3+ \sum_{i\in \partial a} \sigma_i
J_a^i)(3-\sum_{j\in \partial a} \sigma_j J_a^j)}{8}
\label{eq:m12v} \\
& = & \sum\limits_{a=1}^M \frac{9-\sum_{i}\sum_j \sigma_i \sigma_j
J_a^i J_a^j}{8} \\
 &=& \frac{1}{4}\Bigl(3 M + \sum\limits_{i,j}
\sigma_i \mathcal{M}_{i j} \sigma_j
\Bigr) \; ,
\end{eqnarray}
where the matrix element $\mathcal{M}_{i j}$ is defined as
\begin{equation}
\mathcal{M}_{i j} =\left\{
\begin{array}{ll}
- \frac{1}{2} \sum\limits_{a \in \partial i \cap \partial j}
 J_a^i J_a^j  \quad\quad
  & {\rm for} \ \  i\neq j  \; ,\\
0 & {\rm for} \ \  i=j  \; .
\end{array}
\right.
\end{equation}
In the above expressions, $\partial a$ denotes the set of
variables that are connected to clause $a$ by an edge, and
$\partial i$ denotes the set of
clauses that are connected to variable $i$ by an edge, and $\partial i \cap
\partial j$ denotes the intersection of $\partial i$ and $\partial j$.

The maximal eigenvalue of the
symmetric matrix formed by the elements $\mathcal{M}_{i j}$ is denoted
as $\lambda$. This eigenvalue satisfies
\begin{equation}
\lambda \geq \frac{\sum_{i,j} y_i \mathcal{M}_{i j} y_j}{\sum_i y_i^2} \; ,
\label{eq:20110511-01}
\end{equation}
for any non-zero real vector $\underline{y}=(y_1, y_2, \ldots, y_n)$.
Take $y_i=\sigma_i$ for each variable $i$, and it is then easy to show
that $\lambda \geq (4 M_{12} - 3 M)/N$. Combining this with
(\ref{eq20110520-01}), an upper-bound
$M_2^{{\rm upp}}$ for $M_2$ is obtained as
\begin{equation}
\label{eq:M2upp1}
M_2 \leq M_2^{{\rm upp}} \equiv  \frac{1}{2} N \lambda + \frac{1}{2} \sum\limits_{i=1}^{N}
|k_i^+ - k_i^-|  \; .
\end{equation}

If $E_0^{{\rm xor}} > M_2^{{\rm upp}}$ for the given $3$-SAT instance, then the
instance must be unsatisfiable.

\section{Existence of Feige-Kim-Ofek witness for sparse random $3$-SAT}
\label{sec:exist}

Feige and co-authors \cite{Feige-Kim-Ofek-2006} have studied the
existence of FKO witness for random $3$-SAT factor graphs. A random
$3$-SAT factor graph with $N$ variables and $M$ clauses is a
random bipartite graph, with each clause being connected to three randomly
chosen different variables and the edge coupling constant being assigned the
value $+1$ or $-1$ with equal probability.
In the large $N$ limit, it was proved mathematically
in \cite{Feige-Kim-Ofek-2006} that, if the
clause density $\alpha$ grows with $N$ such that
\begin{equation}
\label{eq:FKOscale}
\alpha > c N^{0.4}
\end{equation}
with a sufficiently large constant $c$, then FKO witness exists with
probability approaching $1$ for a random $3$-SAT factor graph of $N$
variables and $\alpha N$ clauses.

However, it is not yet known whether FKO witness exists also for
random $3$-SAT factor graphs with a large but constant clause density
$\alpha$. Here we demonstrate using
the mean-field statistical physics method that, FKO witness should exist
for a random $3$-SAT factor graph with
$\alpha > 19$ in the thermodynamic limit of
 $N\rightarrow \infty$. This estimated constant lower-bound of
clause density is much improved as compared to Eq.~(\ref{eq:FKOscale}).

According to Eq.~(\ref{eq:m12v}), the quantity $M_{1 2}$ can be expressed as
\begin{equation}
M_{1 2} = M - \sum\limits_{a=1}^M \delta
\Bigl( \bigl|
\sum_{j\in \partial a} J_a^j \sigma_j
\bigr| - 3 \Bigr) \; , \label{eq:m12v2}
\end{equation}
where $\delta(x)$ is the Kronecker
symbol, with $\delta(x)=0$ if $x\neq 0$ and
$\delta(x)=1$ if $x=0$.
Combining Eq.~(\ref{eq:m12v2}) with
Eq.~(\ref{eq20110520-01}), we obtain another upper-bound for $M_2$
as
\begin{eqnarray}
M_2^{{\rm max}}  &= & \frac{1}{2} \biggl(M +
\sum_{i=1}^{N} |k_i^+ - k_i^-| \biggr)  \nonumber \\
& & \quad -
2 \min\limits_{\underline{\sigma}} \biggl[\sum\limits_{a=1}^M \delta
\Bigl( \bigl|
\sum_{j\in \partial a} J_a^j \sigma_j
\bigr| - 3 \Bigr)
\biggr] \; .
\label{eq:M2max}
\end{eqnarray}
The first term on the right of Eq.~(\ref{eq:M2max}) is easy to
calculate, while the minimum of the second term over all the
configurations $\underline{\sigma}$ can be evaluated by the
zero-temperature first-step replica-symmetry-breaking (1RSB) cavity method
\cite{Mezard-Parisi-2003,Mezard-Parisi-2001,Mezard-Zecchina-2002,Mezard-Montanari-2009}.
The upper-bound $M_2^{{\rm max}}$
is tighter (smaller) than the
upper-bound $M_2^{{\rm upp}}$ of Eq.~(\ref{eq:M2upp1}).

\begin{figure}
\begin{center}
\includegraphics[width=0.95\linewidth]{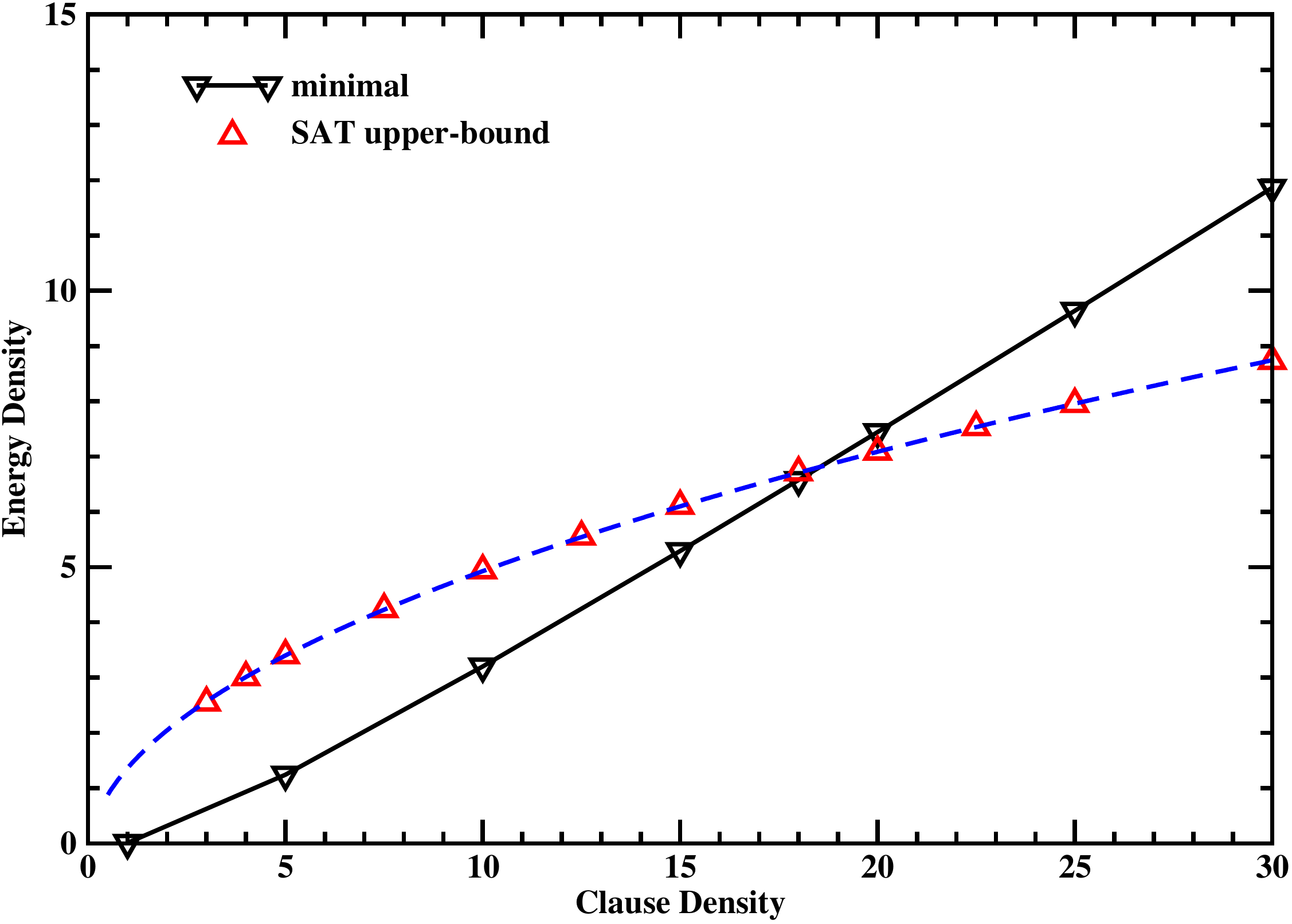}
\end{center}
\caption{
\label{fig:20110520-01}
(color online).
The down-triangles connected by the solid line are
 the minimum energy density $E_0^{{\rm xor}}/N$
of the $3$-XORSAT formula (\ref{eq:eXORSAT}).
The upper-triangles are the upper-bound $M_2^{{\rm max}}/N$
obtained by Eq.(\ref{eq:M2max}) under the assumption that
the $3$-SAT energy (\ref{eq:eSAT}) is satisfiable. The
dashed line is a fitting curve of the form
$M_2^{{\rm max}}/N = c_1 + c_2 \sqrt{\alpha}$. For $\alpha > 19$ the
predicted upper-bound is lower than the global minimum, indicating that the
assumption that Eq.~(\ref{eq:eSAT}) is satisfiable must be wrong.
}
\end{figure}

The global minimum $E_0^{{\rm xor}}$ of the $3$-XORSAT energy (\ref{eq:eXORSAT})
can also be evaluated similarly using the zero-temperature
1RSB cavity method. Figure~\ref{fig:20110520-01} is the comparison
between the value $M_2^{{\rm max}}/N$ and the
ground-state energy density $E_0^{{\rm xor}}/N$ of
(\ref{eq:eXORSAT}) using clause density $\alpha$ as the
control parameter.
When $\alpha > 19$, the requirement that ground-state energy density
$E_0^{{\rm xor}}/N$ being lower than the upper-bound $M_2^{{\rm max}}/N$ is
violated, which gives an indication that the $3$-SAT energy function
(\ref{eq:eSAT}) has no zero-energy configurations. However, when
$\alpha< 19$, $E_0^{{\rm xor}}/N < M_2^{{\rm max}}/N$ is consistent with the
assumption that the $3$-SAT formula is satisfiable, indicating that
no FKO witness exists for the most difficult region of
 $\alpha < 19$.

The random $3$-SAT problem is the hardest when the clause density $\alpha$
is close to the satisfiability threshold $\alpha_s(3)=4.267$
\cite{Mezard-etal-2002,Mezard-Zecchina-2002,Mertens-etal-2006}.
Figure~\ref{fig:20110520-01} suggests that in the hardest UNSAT
region of
$\alpha_s(3) \leq \alpha < 19$ it is impossible to prove $3$-SAT
satisfiability through the FKO witness approach (even if
one can precisely determine the $3$-XORSAT ground-state
energy $E_0^{{\rm xor}}$). When the clause density $\alpha$ of
a random $3$-SAT formula is only slightly
beyond $\alpha_s(3)$, exhaustive enumeration may be the
only way to prove its unsatisfiability.

\section{Witness construction}
\label{sec:sample}

In practice, to find a FKO witness we have to show that the ground-state
energy $E_0^{{\rm xor}}$ of the $3$-XORSAT formula (\ref{eq:eXORSAT}) is
higher  than either $M_2^{{\rm max}}$ or $M_2^{{\rm upp}}$. While the value of
$M_2^{{\rm upp}}$ is easy to calculate, the exact determination of
$E_0^{{\rm xor}}$ is a NP-hard computational problem. Feige and co-authors
tried to circumvent this computational difficulty
by constructing a lower-bound for
$E_{0}^{{\rm xor}}$ \cite{Feige-Kim-Ofek-2006}.  If the value of
this lower bound is higher than $M_2^{{\rm upp}}$, it is guaranteed that
$E_0^{{\rm xor}} > M_2^{{\rm upp}}$.

\subsection{A lower-bound on $E_0^{{\rm xor}}$}

Given a $3$-SAT formula $F$ with $N$ variables and $M$ clauses, a subformula
$f$ is obtained by choosing $m$ clauses from the $M$ clauses.
For such a subformula $f$ its $3$-SAT energy and $3$-XORSAT
energy can be defined similar to Eqs.~(\ref{eq:eSAT}) and (\ref{eq:eXORSAT}).
It is computationally easy to
determine whether a subformula $f$ is $3$-XORSAT satisfiable.

It was noticed in Ref.~\cite{Feige-Kim-Ofek-2006} that,
for a $3$-SAT formula $F$, if $t$ subformulas can be constructed
such that each of them is unsatisfiable as $3$-XORSAT, and each clause
of $F$ appears in at most $d$ of the $t$ subformulas, such that
\begin{equation}
\label{eq:tovd}
\frac{t}{d} >  M_2^{{\rm upp}} \; ,
\end{equation}
then the formula $F$ is unsatisfiable as $3$-SAT.

To prove this statement, we simply notice that, if $F$ is satisfiable as
$3$-SAT, the minimum number of simultaneously unsatisfied clauses as
$3$-XORSAT can not exceed $M_2^{{\rm upp}}$. On the other hand, there are
$t$ unsatisfiable $3$-XORSAT subformulas, meaning that at least $t$ clauses
(some of them might be identical) are simultaneously unsatisfied (as
$3$-XORSAT) by any spin configuration.
Since each clause can be present in at most $d$ different subformulas,
the total number of simultaneously unsatisfied different clauses is at least
$t/d$ \cite{Feige-Kim-Ofek-2006}.

Let us point out a simple improvement over the criterion Eq.~(\ref{eq:tovd}).
Suppose we have a set of $t$
unsatisfiable $3$-XORSAT subformulas constructed from
the $3$-SAT formula $F$. Let us denote by $d_a$ the number of times
clause $a$ appear in these subformulas. Let us rank the $M$ values of
$d_a$ in descending order and denote the ordered
values as $\{d^{(1)}, d^{(2)},
\ldots, d^{(M)}\}$, with $d^{(1)} \geq d^{(2)} \geq \ldots \geq d^{(M)}$.
A better refutation inequality can be written as
\begin{equation}
C >  M_2^{{\rm upp}} \;  ,
\label{eq:20110520-03}
\end{equation}
where $C$ is the minimal integer satisfying
\begin{equation}
\sum\limits_{a=1}^{C} d^{(a)} \geq t  \; .
\label{20110520-04}
\end{equation}
To prove that (\ref{eq:20110520-03}) ensures
the unsatisfiability of the $3$-SAT formula $F$,
 we only need to show that the ground-state energy
$E_0^{{\rm xor}}$ of the $3$-XORSAT energy (\ref{eq:eXORSAT})
can not be lower than $C$. We reason as follows. To
make $F$ satisfiable as $3$-XORSAT, some clauses have to be removed from
$F$ in such a way that for each of the $t$ constructed unsatisfiable
subformulas, at least one of the involved clauses should be removed.
Therefore, the sum of numbers $d_a$ of the removed clauses
should be at least $t$.
This then proves the refutation inequality (\ref{eq:20110520-03}).
The quantity $C$ as obtained by Eq.~(\ref{20110520-04}) is a lower-bound
of $E_0^{{\rm xor}}$.
This lower-bound actually is not tight, it is much lower than the
true ground-state energy.

\subsection{Random sampling}

A simple way of constructing unsatisfiable $3$-XORSAT witnesses for a
given $3$-SAT formula $F$ are the
following:

\begin{description}

\item[0.] Calculate $\sum_i |k_i^+ - k_i^- |$ and the maximal eigenvalue
$\lambda$ of matrix $\mathcal{M}$ for formula
$F$. Set subformula number as $t=0$ and set
the counting number $d_a=0$ for each clause $a$
of $F$.

\item[1.] Randomly select $N^\gamma$ variables from the set of $N$ variables,
where $\gamma \in [0,1]$ is a fixed parameter.

\item[2.] Check if the subformula $f$ of $F$ induced by these $N^\gamma$
variables is $3$-XORSAT satisfiable, and if yes, go back to step {\bf 1}.
Otherwise a unsatisfiable $3$-XORSAT formula is obtained.

\item[3.] Construct a subformula $\tilde{f}$
by adding clauses of $f$ one after the other in a random order, until
$\tilde{f}$ becomes unsatisfiable (and has ground-state energy $1$)
as $3$-XORSAT. Then prune the subformula $\tilde{f}$
 by recursively removing those variables
that are connected to only one clause and the associated single clauses.
After this leaf-removal process is finished, we obtain an unsatisfiable
core subformula. The counting number $d_a$ of each clause of this core
subformula is increased by one ($d_a \leftarrow d_a + 1$), and the subformula
number is also increased by one ($t \leftarrow t+1$).

\item[4.] Calculate $C$ according to (\ref{20110520-04}) and then check if
(\ref{eq:20110520-03}) is satisfied. If yes, output $`$UNSAT witness found';
 otherwise repeat steps {\bf 1}-{\bf 4}.
\end{description}

\begin{figure}
\begin{center}
\includegraphics[width=0.95\linewidth]{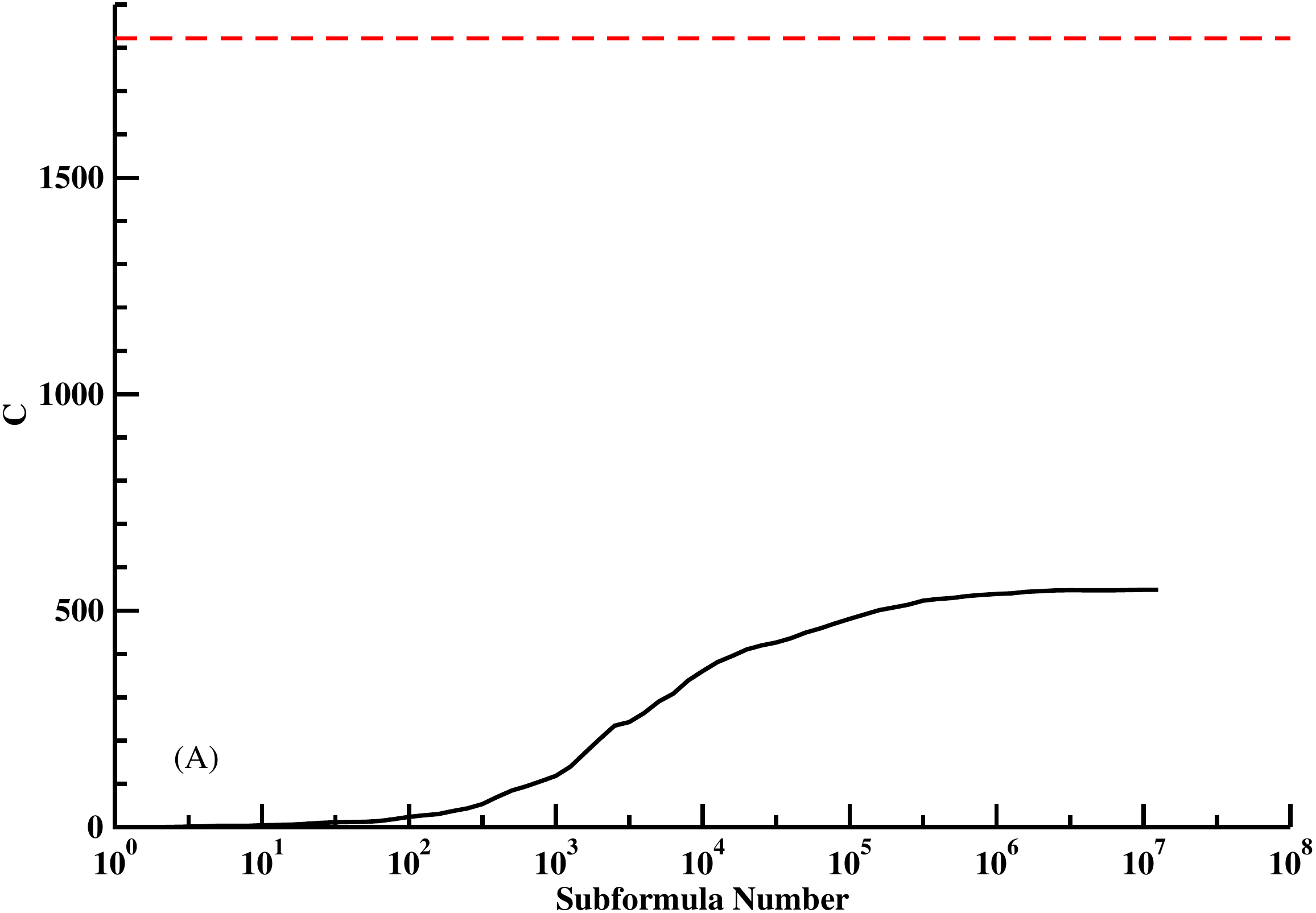}
\vskip 0.6cm
\includegraphics[width=0.95\linewidth]{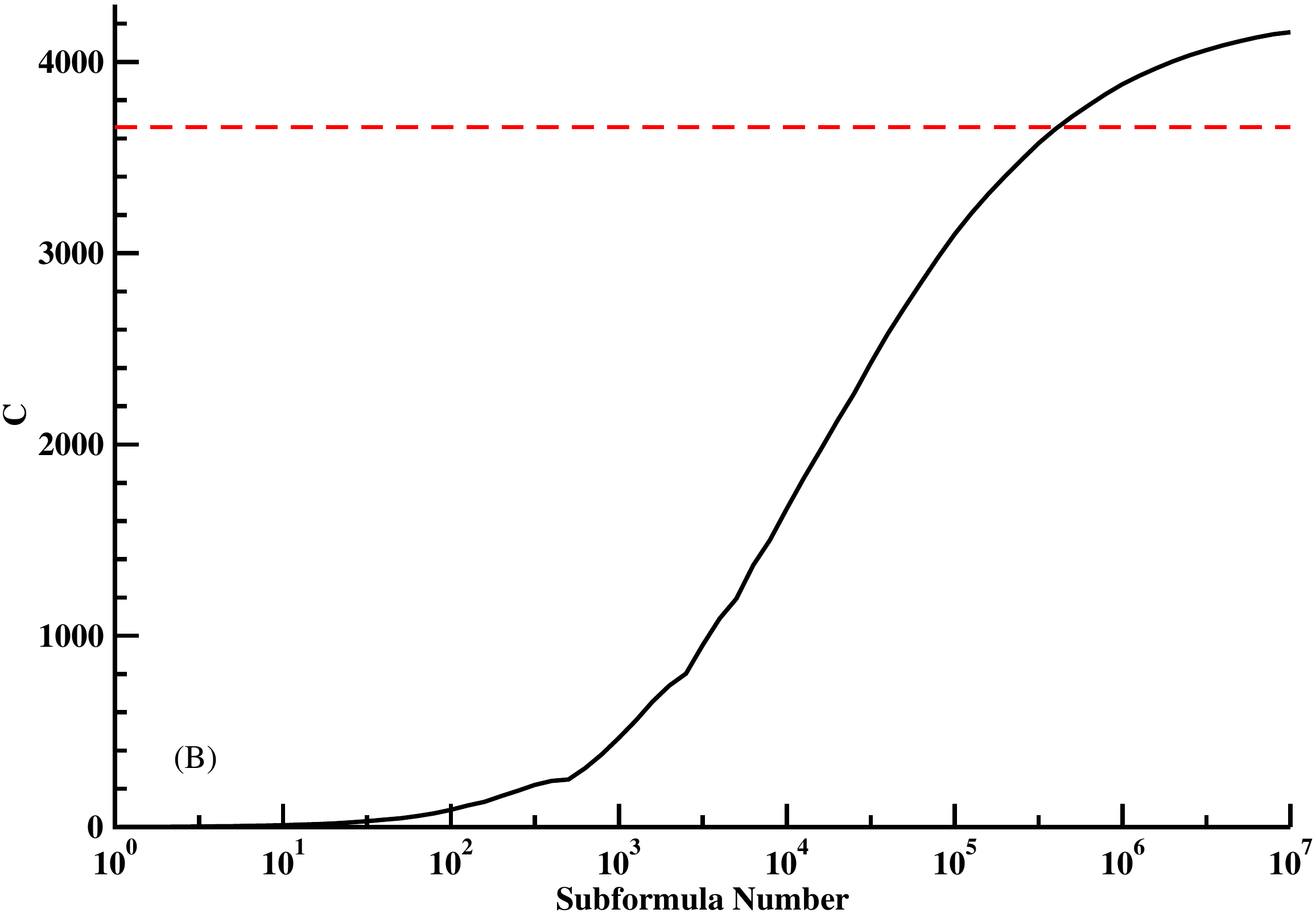}
\end{center}
\caption{
\label{fig:evolutionC}
(color online).
The evolution of witness value $C$ as the number of
randomly sampled subformulas $t$  increases. The investigated
random $3$-SAT instance has variable number $N=100$ and
clause number $M=10,000$ in (A) and $M=40,000$ in
(B). The horizontal dashed lines in (A) and (B)
mark the position of $M_2^{{\rm upp}}$.
The control parameter $\gamma$ of the random sampling algorithm is set to
$\gamma=0.5$.}
\end{figure}

Figure \ref{fig:evolutionC} shows the simulation results on two single
$3$-SAT instances. The upper panel $A$ is a $3$-SAT formula with $100$ variables
and clause density $\alpha=100$, and the lower panel $B$ is another
$3$-SAT formula with $100$ variables and clause density $\alpha=400$.
If the curve $C(t)$ is able to go beyond $M_2^{{\rm upp}}$ (marked by the
horizontal dashed line) then a FKO witness is found.
The random sampling algorithm succeeded in finding a FKO witness for
the instance with $\alpha=400$ but failed to do so for the
one with $\alpha=100$.

For $N\gg 1$, a random subformula as constructed by the above-mentioned
procedure contains about $0.633 N^\gamma$ clauses \cite{Mezard-etal-2003}.
When there are a large number $t$ of such subformulas, the total number of
clauses is about $0.633 t N^\gamma$, and each clause appears on average
in $\overline{d}=0.633 t N^\gamma /M$ subformulas.
From this we estimate that the solution
$C$ of (\ref{20110520-04}) is roughly
\begin{equation}
C \sim \frac{t}{\overline{d}} \approx  \frac{M}{N^\gamma} =
\alpha N^{1-\gamma} .
\end{equation}
On the other hand, $M_2^{{\rm upp}}$ scales as $\alpha^{1/2} N$
(see Fig.~\ref{fig:20110520-01} and
\cite{Feige-Kim-Ofek-2006}). Therefore, we see that for
the inequality (\ref{eq:20110520-03})
to hold, it is required that
\begin{equation}
\alpha > N^{2 \gamma} .
\label{20110520-06}
\end{equation}
The average number of clauses among a randomly chosen $N^\gamma$ variables
is about $N^{3 \gamma -3} M = \alpha N^{3\gamma - 2}$. This value should be
proportional to $N^\gamma$ so that the subformula induced by these
variables has a high probability to be unsatisfiable as $3$-XORSAT. Therefore
we require that $\alpha N^{3\gamma-2} \approx N^\gamma$, from which we get
\begin{equation}
\alpha \approx N^{2-2 \gamma} .
\label{20110520-07}
\end{equation}
From Eqs.~(\ref{20110520-06}) and (\ref{20110520-07}) we obtain that
the parameter $\gamma$ should be chosen as
\begin{equation}
\gamma = \frac{1}{2} .
\end{equation}

The above analysis suggests that, for random $3$-SAT instances with clause
density $\alpha > N$, it is relatively easy to construct UNSAT witnesses.
However,
for clause density sublinear in $N$, it is very hard to construct UNSAT
witnesses through the above random process.

\begin{figure}
\begin{center}
\includegraphics[width=0.9\linewidth]{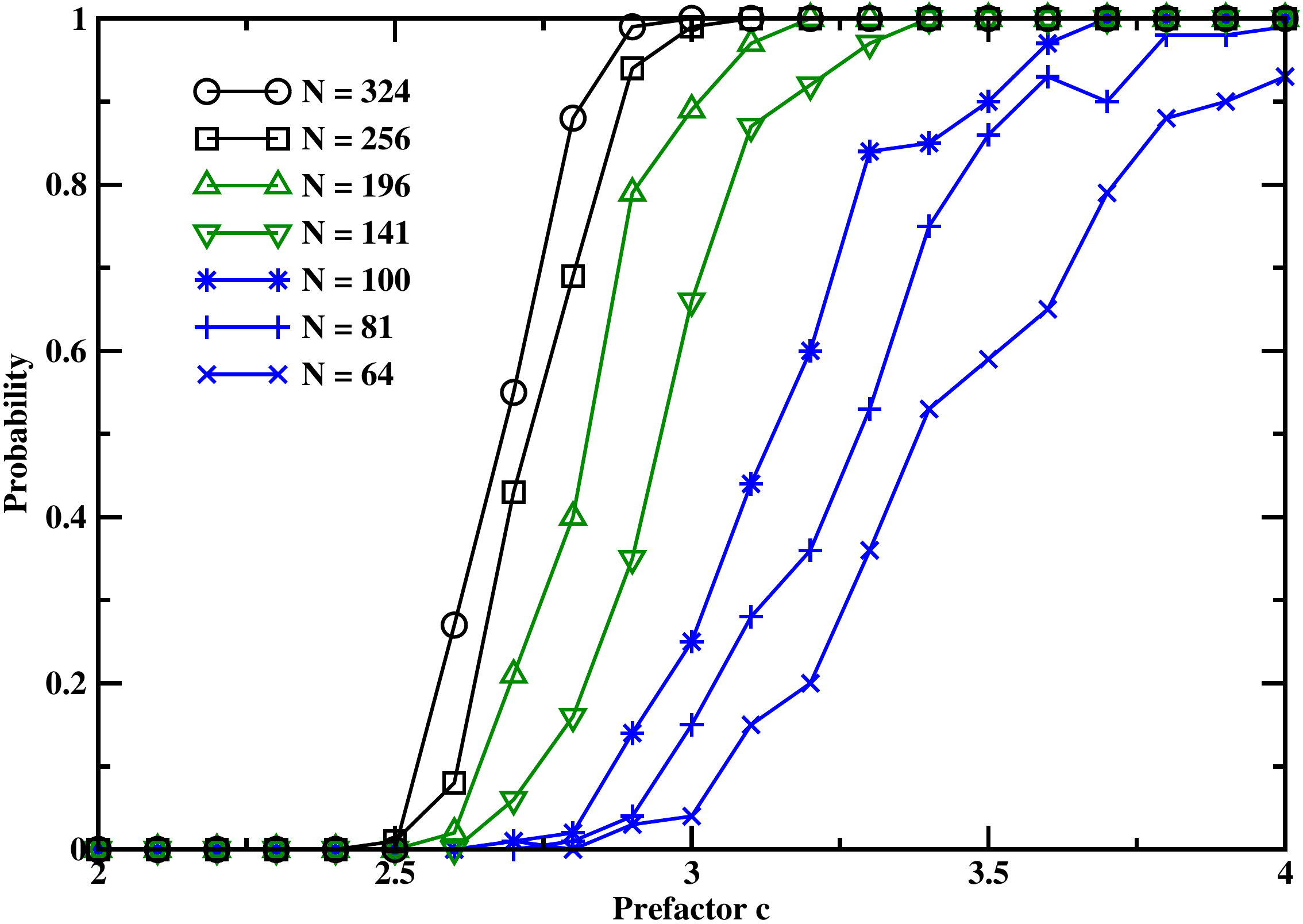}
\end{center}
\caption{\label{fig:20110520-08}
(color online).
The probability of FKO witness being found in a single run of the random
sampling process.  Each data point was obtained by simulating $10$
 random $3$-SAT instances with $N$ variables and $M= c N^2$ clauses.
 Different curves correspond to
different variable numbers $N$.
}
\end{figure}

The performance of this random construction process, with $\gamma=0.5$, is
demonstrated in Fig.~\ref{fig:20110520-08} for random $3$-SAT formulas with
clause density $\alpha = c  N$. This figure shows that for clause density
scales linearly with variable number $N$, the prefactor $c$ needs to be
greater than $c \approx 2.5$ for the random sampling algorithm to find
FKO witnesses.

\begin{figure}
\begin{center}
\includegraphics[width=0.9\linewidth]{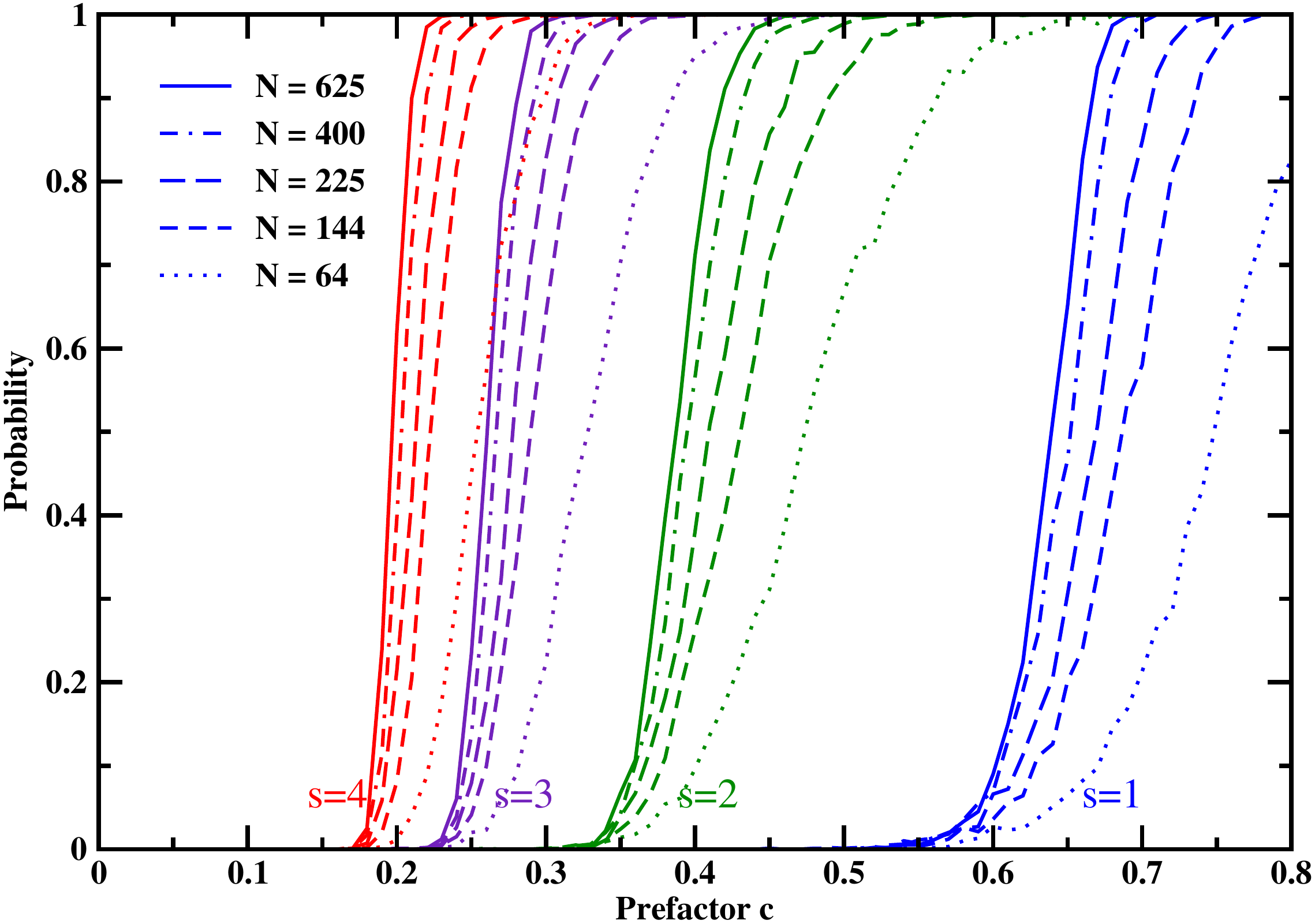}
\end{center}
\caption{\label{fig:satwitness}
(color online).
The probability of a random $3$-SAT formula with $N$ variables and $c N^2$ clauses
(clause density $\alpha = c N$) being proven to be UNSAT by the $2$-SAT
refutation algorithm. The
seed size $s$ is fixed to $s=1$, $s=2$, $s=3$, and $s=4$ in the four sets of simulation curves. Each curve is the average over $10$ random instances.
}
\end{figure}

The random sampling algorithm is therefore
very inefficient in obtaining FKO witnesses. For clause density $\alpha$
linear in $N$ other local refutation algorithms are more efficient.
For example, a simple $2$-SAT refutation algorithm goes as follows.
 First, a seed set of size $s$ is chosen, which contains the $s$ variables of the
highest degrees. Each of the $2^s$ spin assignments of these $s$ variables
will induce a $2$-SAT subformula, and we can check whether this $2$-SAT subformula
is satisfiable or not. If all these $2^s$ induced $2$-SAT subformulas are UNSAT,
then the original $3$-SAT formula can not be satisfied. The number of clauses
in the induced $2$-SAT subformula is about $ \frac{3}{2} s \alpha$, and the number of
variables is at most $N$. Since a random $2$-SAT formula is very likely to be
unsatisfiable if the number of clauses exceeds the number of variables, then we
see that the simple $2$-SAT refutation algorithm has a high probability of
success if $\alpha > \frac{2}{3 s} N$. The simulation results shown in
Fig.~\ref{fig:satwitness} confirm this expectation.

\subsection{Focused local search}

The subformulas constructed by the random sampling algorithm are very sparse. Most of
 the loops in such a subformula are long-ranged, with lengths scaling logarithmicly
 with the number of variables. We now consider another construction strategy,
 namely focused local search.
The goal of this strategy is to construct $3$-XORSAT unsatisfiable subformulae
with only short loops.

The details of the focused local search algorithm are as follows:
\begin{enumerate}
\item[0.] The used set $U$ of clauses is initialized as empty.

\item[1.] Arbitrarily choose a clause $a$ that does not belong to the set $U$. This clause and all its attached three vertices form the $``$system", $I$.  Any clause $b$ that is connected to the $``$system" by at least one edge and is not
    in $U$ belongs to the $``$boundary", $B$.

\item[2.] In the $``$boundary" $B$ some of the clauses have more connections to the
$``$system" than the other clauses. Randomly choose a clause $c$ in the
$``$boundary" that has the maximal number of connections with the
$``$system" (i.e., the number of edges to the $``$system" is the
maximal among all the clauses in the $``$boundary").
Include clause $c$ and all its attached vertices to the $``$system", and
add clause $c$ to the set $U$.
The $``$boundary" $B$ is then updated. Clause $c$ is removed from $B$,  all the clauses that are connected to the $``$system" and that are not belong to the set $U$ are added to $B$.

\item[3.] Check whether the $``$system" is $3$-XORSAT satisfiable, if yes and the $``$boundary" $B$ is
    not empty, go back to step {\bf 2}. If the $``$system" is $3$-XORSAT unsatisfiable,
    then go to step {\bf 4}. If the
    $``$system" is still satisfiable but the boundary $B$ becomes empty,
    then stop and output $`$construction failed'.

\item[4.] After an unsatisfiable $3$-XORSAT subformula is obtained, the number
of unsatisfied clauses in this subformula is $1$. We then prune the subformula
 by removing unnecessary clauses so that  an unsatisfiable core subformula is obtained.
 In the pruning process, basically we test (in a random order) whether each clause can be removed from the subformula without making it $3$-XORSAT satisfiable.
  If a clause is removed from
 the subformula it is also removed from the used clause set $U$.

\item[5.] Update the subformula number $t$ to $t+1$. If $t\leq M_2^{{\rm upp}}$, go back to
step {\bf 1}, otherwise stop and output $`$UNSAT witness found'.
\end{enumerate}
\begin{figure}
\begin{center}
\includegraphics[width=0.9\linewidth]{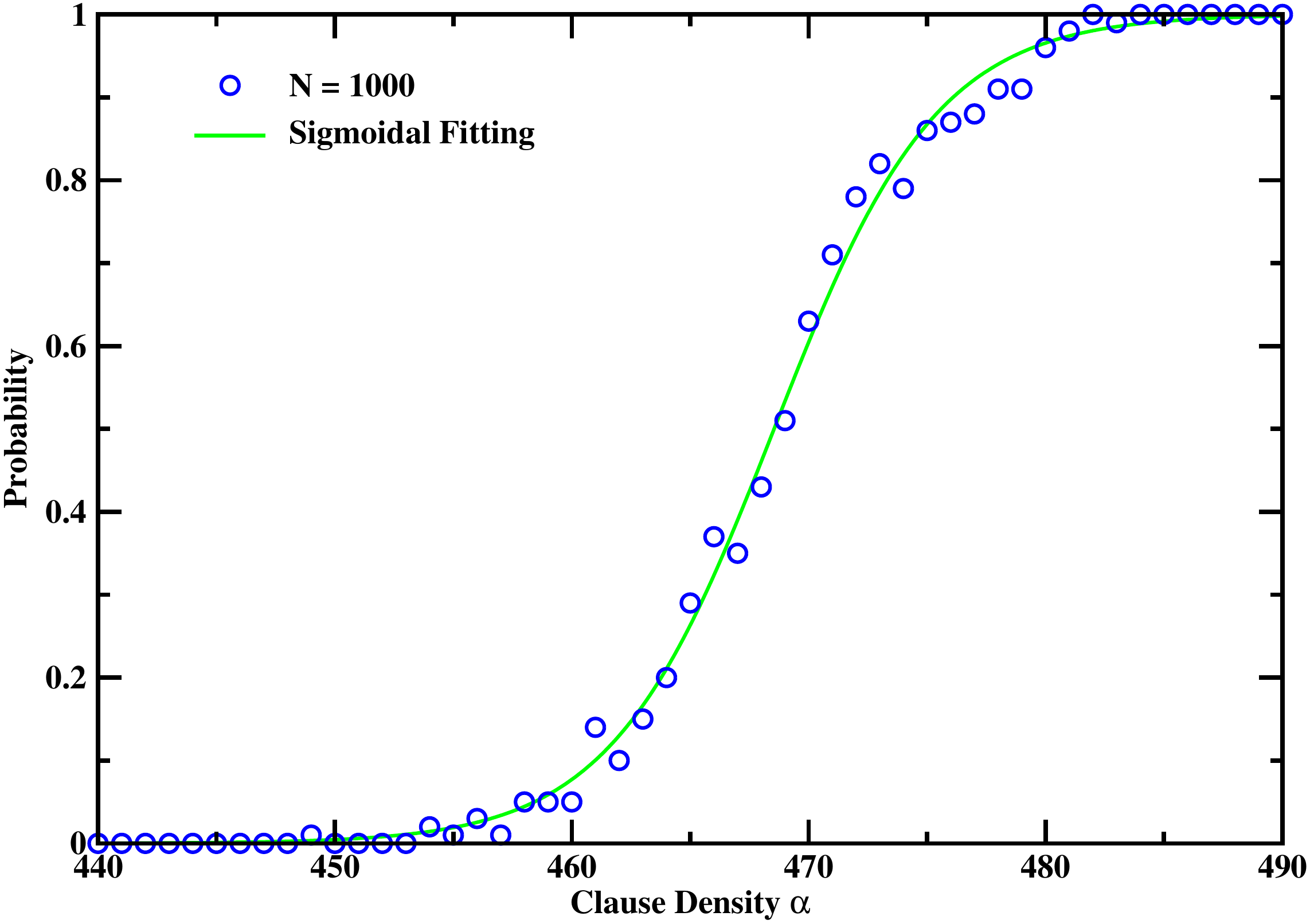}
\end{center}
\caption{\label{fig:flc}
(color online).
The probability of FKO witness being found in a single run
of the focused local search process (control parameter $S=1$)
for random $3$-SAT instances
with $N=1000$ variables and $M = \alpha N$ clauses. Each data
point was obtained by simulating $100$ random $3$-SAT instances.
The solid line is a sigmoidal fitting curve with parameters
$\alpha_0=468.54 \pm 0.09$ and $\Delta = 3.44 \pm 0.08$.
}
\end{figure}

\begin{figure}
\begin{center}
\includegraphics[width=0.9\linewidth]{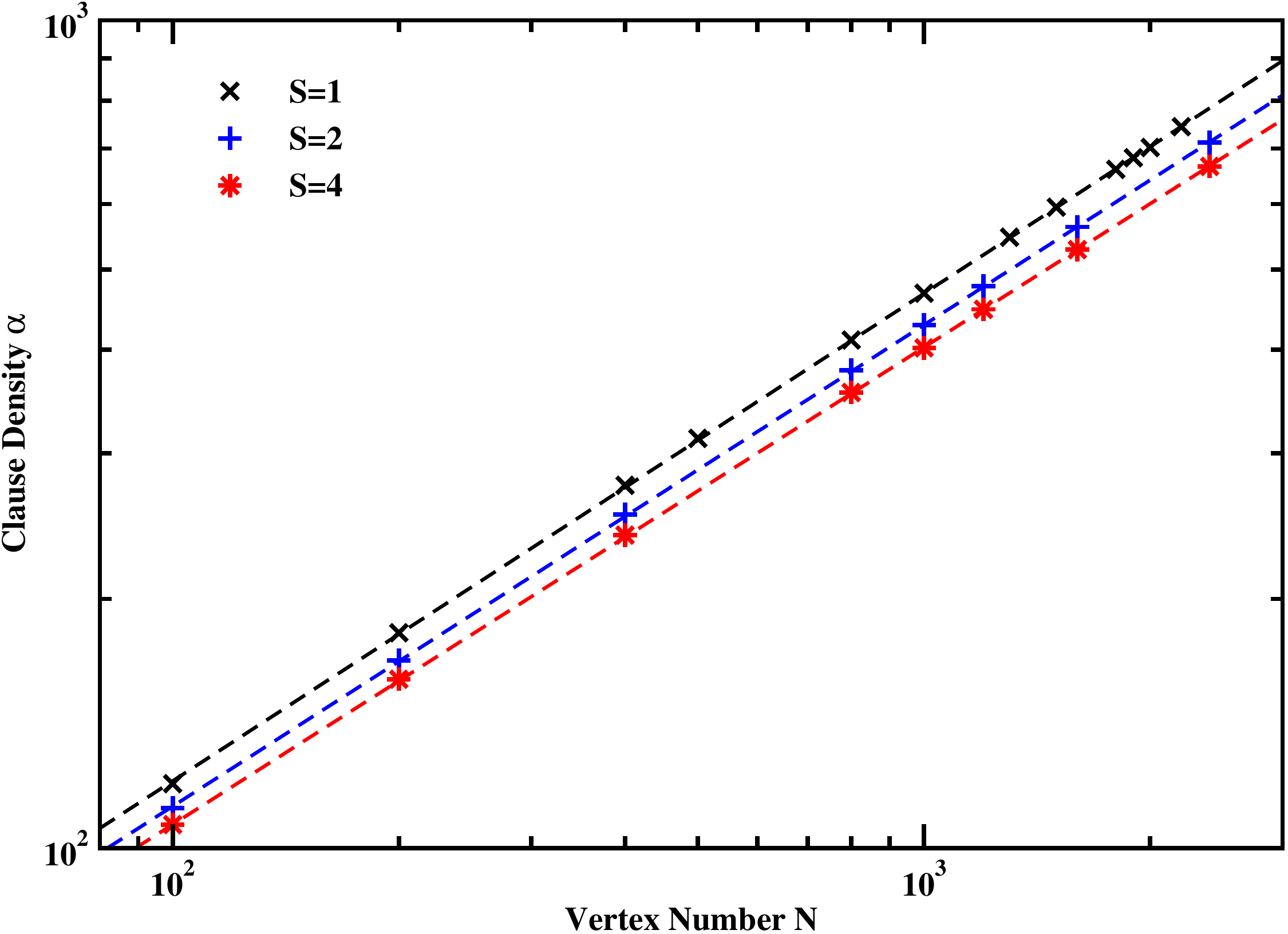}
\end{center}
\caption{\label{fig:ScalingFocus}
(color online).
Scaling behavior between variable number $N$ and the characteristic clause
density $\alpha=\alpha_0$ of the
focused local search algorithm. The control parameter of the focused local
search algorithm is $S$.
The dashed lines are fitting curves of the form $\alpha_0 = c \times N^b$. The
fitting parameters are $c=8.0 \pm 0.1$ and $b=0.589 \pm 0.002$ (top, $S=1$);
$c=7.7 \pm 0.1$ and $b=0.582 \pm 0.002$ (middle, $S=2$); and
$c=7.5 \pm 0.1$ and $b=0.577 \pm 0.002$ (bottom, $S=4$).
}
\end{figure}

In the above-mentioned focused local search algorithm, each clause
can only appear in $S=1$ subformula. Therefore all the constructed
subformulas are disjoint in the sense that they do not share
any clauses. Figure~\ref{fig:flc} shows the performance of this focused
local search algorithm on a set of random $3$-SAT instances with $N=1000$
variables. As the clause density $\alpha$ increases around certain threshold
value $\alpha_0$, the probability of
finding a FKO witness increases quickly from $0$ to $1$.
The simulation data can be well fitted by
a sigmoidal curve
\begin{equation}
P(\alpha) = \frac{1}{1+\exp\bigl(-\frac{\alpha-\alpha_0}{\Delta}\bigr)} \; ,
\end{equation}
where the parameter $\Delta$ controls the slope of the sigmoidal curve.
At $\alpha=\alpha_0$ the focused local search algorithm has $1/2$ probability of
successfully constructing a FKO witness for a random $3$-SAT instance of
$N$ variables. We therefore take $\alpha_0$ as a quantitative measure of
the algorithmic performance. The scaling of $\alpha_0$ with variable
number $N$ is shown in
Fig.~\ref{fig:ScalingFocus}. We find that
\begin{equation}
\alpha_0 \approx c \times N^{b} \; ,
\end{equation}
with exponent $b\approx 0.589$ and prefactor $c \approx 8.0$.
The exponent $b$ is much larger than the value of $0.4$, which was
predicted to be achievable at least by a weak exponential-complexity
algorithm \cite{Feige-Kim-Ofek-2006}. It is also larger 
than the value of $0.5$
achieved by the spectral methods \cite{Goerdt-Krivelevich-2001,Feige-Ofek-2004,CojaOghlan-Goerdt-Lanka-2007}.
At the moment we do not have any analytical argument as regards the value of $b$ of
the focused local search algorithm.

We find that, if we allow each clause to be present in $S\geq 2$ subformulas,
the performance of the focused local search algorithm will be improved. The
scaling behaviors of this modified algorithm with $S=2$ and $S=4$  are
also shown in Fig.~\ref{fig:ScalingFocus}. The simulation data
suggest that both the scaling exponent $b$ and the
prefactor $c$ decrease slightly with $S$. As we have not yet performed
systematic simulations for large values of $S$, we do not know
to what extent the exponent $b$ can be reduced.

\section{Conclusion and Discussions}
\label{sec:further}

In this paper, we demonstrated through mean-field calculations that
a type of unsatisfiability witness, the Feige-Kim-Ofek witnesses,
exists in the random $3$-SAT problem with constant
clause density $\alpha > 19$. However for $\alpha < 19$ our
theoretical result concludes that it is \emph{impossible} to
refute a random $3$-SAT formula through the FKO approach.
We investigated the empirical performances of two witness-searching
algorithms by computer simulations.
The naive random sampling algorithm is able to construct FKO witnesses only
for random $3$-SAT instances with clause density $\alpha > c N$ (where $N$ is the
variable number). The focused local search algorithm has much better
performances, it works for $\alpha > c N^b$ with $b\approx 0.59$.
The value of the exponent $b$ can be further decreased by enlarging the
control parameter $S$ of the focused local search algorithm.
It would be interesting to systematically investigate the relationship between
$b$ and $S$ by computer simulations in a future work.

The essence of the FKO witness is to construct a rigorous lower-bound for the
ground-state energy $E_0^{{\rm xor}}$ of the $3$-XORSAT formula (\ref{eq:eXORSAT}).
The tighter this lower-bound to $E_0^{{\rm xor}}$ is, the better the refutation power
of this witness approach.
A very big theoretical and algorithmic challenge is to
obtain a good lower-bound for the ground-state energy of
the $3$-XORSAT problem.
For the $3$-SAT problem, H{\aa}stad proved in Ref.~\cite{Hastad-1997} that
no algorithm is guaranteed to construct spin assignments that can satisfy
more than $(7/8) M_{opt}$ clauses in polynomial time ($M_{opt}$ being
the maximal number clauses that can be simultaneously satisfied),
unless $P=NP$. This actually gives an upper bound on the ground-state
energy of the $3$-SAT problem. This upper-bound can be converted to an upper-bound
for $E_0^{{\rm xor}}$ of the $3$-XORSAT problem. But we do not know any
energy lower-bound for the $3$-XORSAT problem whose value is proportional to the
clause density $\alpha$. If such an energy lower-bound can be verified algorithmically,
then the FKO witness approach will succeed for the $3$-SAT problem with constant
$\alpha$.

The $3$-XORSAT energy lower bound $C$ as obtained from Eq.~(\ref{20110520-04})
does not scale linearly with the clause density $\alpha$ but only
sublinearly. One possible way of improving the value of $C$ goes as follows.
For each
constructed $3$-XORSAT unsatisfiable subformula $f$, we assign a properly chosen real-valued weight $w_f$. Correspondingly the counting number $d_a$ of each clause
$a$ is modifed as
\begin{equation}
d_a = \sum\limits_{\{f | a \in f\}} w_f \; ,
\end{equation}
where the summation is over all the subformulas $f$ that contain clause $a$.
Then Eq.~(\ref{20110520-04}) is changed into
\begin{equation}
\label{eq:newC}
\sum\limits_{a=1}^{C} d^{(a)} \geq \sum_{f} w_f \; .
\end{equation}
When all the weights $w_f=1$, then Eq.~(\ref{eq:newC}) reduces to
Eq.~(\ref{20110520-04}). 
By optimizing the choices of the subformula weights $\{w_f\}$ we
expect that a considerly better energy lower bound $C$ can be obtained
from Eq.~(\ref{eq:newC}). 

The counting number $d_a$ of each clause $a$ can also be considered as a
real-valued parameter whose value can be freely adjusted. Then the weight
of each constructed subformula $f$ is defined as
$w_f=\min\limits_{a \in f} d_a$ (i.e., the lowest value of $d_a$ over all
the clauses of $f$).  We believe another better energy lower bound $C$ can
also be obtained by optimizing the choices of $\{d_a\}$.

A systematic exploration of these two re-weighting schemes and other 
 possible extensions will be carried out in a separate study.

\section*{Acknowledgments}

This work was initialized during the KITPC program $``$Interdisciplinary Applications of
Statistical Physics and Complex Networks" (Beijing, 2011).
We thank Elitza Maneva and Osamu Watanabe for suggesting this interesting
problem to us,
and Andr\'{e} Medieros for his initial involvement in this project. LLW thanks
Chuang Wang and Ying Zeng for helps on programming.
This work was supported by the Knowledge Innovation Program of Chinese
Academy of Sciences (No.~KJCX2-EW-J02) and the National Science Foundation of China
(grant No.~11121403 and 11225526). It was also supported by the
Academy of Finland as part of its Finland Distinguished
Professor program, Project No. 129024/Aurell and through
the Centres of Excellence COIN (Aurell) and COMP (Alava).


\end{document}